\documentclass[10pt,conference]{IEEEtran}
\usepackage[pdftex]{graphicx}

\usepackage{subfig}

\usepackage{algorithm}
\usepackage{algorithmic,xpatch}
\usepackage{subfloat}
\usepackage{array}
\usepackage{verbatim}

\IEEEoverridecommandlockouts
\IEEEpubid{%
   \makebox[\columnwidth]{ISBN 978-3-901882-94-4~\copyright~2017 IFIP \hfill}%
   \hspace{\columnsep}%
   \makebox[\columnwidth]{ }%
}

\begin{document}

\title{Observing IoT Resources over ICN}

\author{\IEEEauthorblockN{Hasan M A Islam, Dmitrij Lagutin}
\IEEEauthorblockA{Department of Computer Science\\
Aalto University\\
Espoo, Finland\\
Email: firstname.lastname@aalto.fi}
\and
\IEEEauthorblockN{Nikos Fotiou}
\IEEEauthorblockA{Department of Informatics\\
Athens University of Economics and Business\\
Athens, Greece\\
Email: fotiou@aueb.gr}}

\maketitle

\begin{abstract}
The Constrained Application Protocol (CoAP) is an HTTP-like protocol for RESTful applications intended to run on constrained devices, typically part of the Internet of Things. CoAP observe is an extension to the CoAP specification that allows CoAP clients to observe a resource through a simple publish/subscribe mechanism. In this paper we leverage Information-Centric Networking (ICN), transparently deployed within the domain of a network provider, to provide enhanced CoAP services. We present the design and the implementation of CoAP observe over ICN and we discuss how ICN can provide benefits to both network providers and CoAP applications, even though the latter are not aware of the existence of ICN. In particular, the use of ICN results in smaller state management and simpler implementation at CoAP endpoints, and less communication overhead in the network.      
\end{abstract}

\IEEEpeerreviewmaketitle

\section{Introduction}
The Internet of Things (IoT) is expected to interconnect billions of devices that will generate a vast amount of information. Significant research efforts have been devoted into enabling smart devices to connect to the Internet, share information, and consume services. These efforts have resulted in a variety of network access technologies and higher layer protocols. On the other hand, core (inter-)networking technologies have not been adapted to this new paradigm, raising concerns about whether or not networks will be able to cope with the scale and the patterns of the traffic of the IoT. In order to assuage these concerns a number of researches have sprung up proposing Future Internet (FI) architectures. One such promising FI architecture is Information-Centric Networking (ICN). \footnote{A survey of ICN research and architectures that have been investigated, with some still being pursued and experimentally explored further, can be found in~\cite{Xyl2013}.} ICN advocates implementing all (inter-)networking functions around content (i.e., information) identifiers, rather than location identifiers. This shift in focus and techniques is expected to overcome various limitations of the current Internet~\cite{Tro2010}. However, such a shift requires not only the re-design of networking protocols, but also the modification of legacy Internet applications. Such radical changes at all network layers are an overwhelming barrier to the adoption of the ICN. With this in mind, the POINT project~\cite{Tro2015} proposes a radical approach to ICN adoption: it postulates an individual ICN operator that uses network attachment points to translate \emph{legacy} IP applications traffic to ICN, i.e., the endpoints are oblivious to the ICN.

The Constrained RESTful Environments (CoRE) working group has designed and developed the Constraint Application Protocol (CoAP)~\cite{rfc7252} which is intended to operate in the constrained IP networks and provides the RESTful services to constrained devices. The CoAP interaction model is very similar to the client/server model of HTTP: a CoAP client issues a request message to a server and if the CoAP server is able to serve the request, it responds with a response code and the payload to the requester. Unlike HTTP, CoAP requests and responses are exchanged asynchronously on top of an unreliable datagram oriented transport protocol (e.g., UDP). 

CoAP observe, described in \cite{rfc7641}, is an extension to the CoAP specification. The CoAP observe enables a CoAP client to observe a resource hosted in a IoT device through a simple publish/subscribe mechanism. The CoAP client registers with the CoAP server for a particular resource. If the server accepts the registration, it asynchronously pushes notifications of the resource state changes to the interested clients and follows a best-effort approach to guarantee the eventual consistency of the observed state and the actual state of the resource. Compared to HTTP, the CoAP observe can significantly reduce the communication overhead in terms of the bandwidth requirements and the number of messages transmitted. Since the CoAP observe protocol is based on the publish/subscribe paradigm, it can benefit from the POINT architecture in terms of latency, state management, communication overhead, and better security and privacy.

In this paper, we present the design and implementation of the CoAP observe over ICN and describe how CoAP can benefit from the POINT architecture as well as how a network operator that offers the CoAP connectivity can benefit from ICN by leveraging the multicast capabilities of the POINT architecture. It is important to note that our solution enables the usage of \textit{legacy IP-based devices}, for instance, the existing CoAP endpoints can transparently access the resources hosted in IoT devices through the intermediate ICN network.

The remainder of this paper is organized as follows. In Section~\ref{sec:back} we introduce the CoAP and the CoAP observer protocols, as well as, the POINT architecture. In Section~\ref{sec:coap} we illustrate CoAP observe over ICN reference architecture, alongside with the implementation details, and we highlight how ICN can benefit CoAP-based applications. In Section~\ref{sec:node} we present our CoAP observe over ICN design and implementation. In Section~\ref{sec:related} presents related work in the area, and finally Section~\ref{sec:conc} concludes our paper.

\section{Background}
\label{sec:back}
\subsection{CoAP and CoAP Observe}
 
CoAP~\cite{rfc7252} has been designed and developed to be a 'lightweight HTTP' so that it can be suitable to operate in the constrained IP networks. The CoAP interaction model is similar to the client/server model of HTTP: a CoAP client issues a request message to a server and if the CoAP server is able to serve the request, it responds to the requester with a response code and the payload. Unlike HTTP, CoAP requests and responses are exchanged asynchronously on top of an unreliable datagram oriented transport protocol (e.g., UDP). The CoAP messaging model supports 4 types of messages: CON (confirmable), NON (non-confirmable), ACK (Acknowledgement), RST (Reset). Every CoAP message carries a Token whose value is a sequence of 0 to 8 bytes. The Token correlates a response with a request, along with the additional address information of the corresponding CoAP endpoint. The CoAP client generates a Token for a request message and the server uses the same Token in the response. Each message also contains a 16-bit message ID, which is used to detect message duplicates.

The CoAP protocol also supports intermediaries and caching of responses. There are two different kinds of proxies: Forward-Proxy and Reverse-Proxy. A Forward-Proxy sends a CoAP request to the CoAP server on behalf of a CoAP client. For this, the Forward-Proxy needs to be configured to perform requests on behalf of the client. In contrast, the Reverse-Proxy is transparent to the client. The Reverse-Proxy behaves as if it were the server of origin. The CoAP protocol supports the discovery of resources by exploiting a separate entity called Resource Directory (RD) which stores the descriptions of resources. Moreover, CoAP supports group communication~\cite{rfc7390} based on IP multicast; CoAP groups and the membership of a group can be discovered via the lookup interfaces in the Resource Directory (RD). Finally, CoAP enables clients to observe a resource through a simple publish/subscribe mechanism ~\cite{rfc7641}. With this, the server asynchronously pushes the notification of state changes of the resource for which the client is interested in and follows a best-effort approach to guarantee the eventual consistency of the observed state and the actual state of the resource.  

The CoAP observe protocol supports resource observation; for this to be done, a CoAP client needs to register with a CoAP server using a GET request with the observe option setting the value to 1 (one). To unsubscribe the resource observation, the CoAP client sends ACK by setting the value of the CoAP observe 0 (zero). The CoAP client can also utilize a proxy for observing a resource. Then, everytime the state of the observed resource changes, the server pushes the notification back to the client(s).

\begin{figure}
\includegraphics[height=.7\columnwidth]{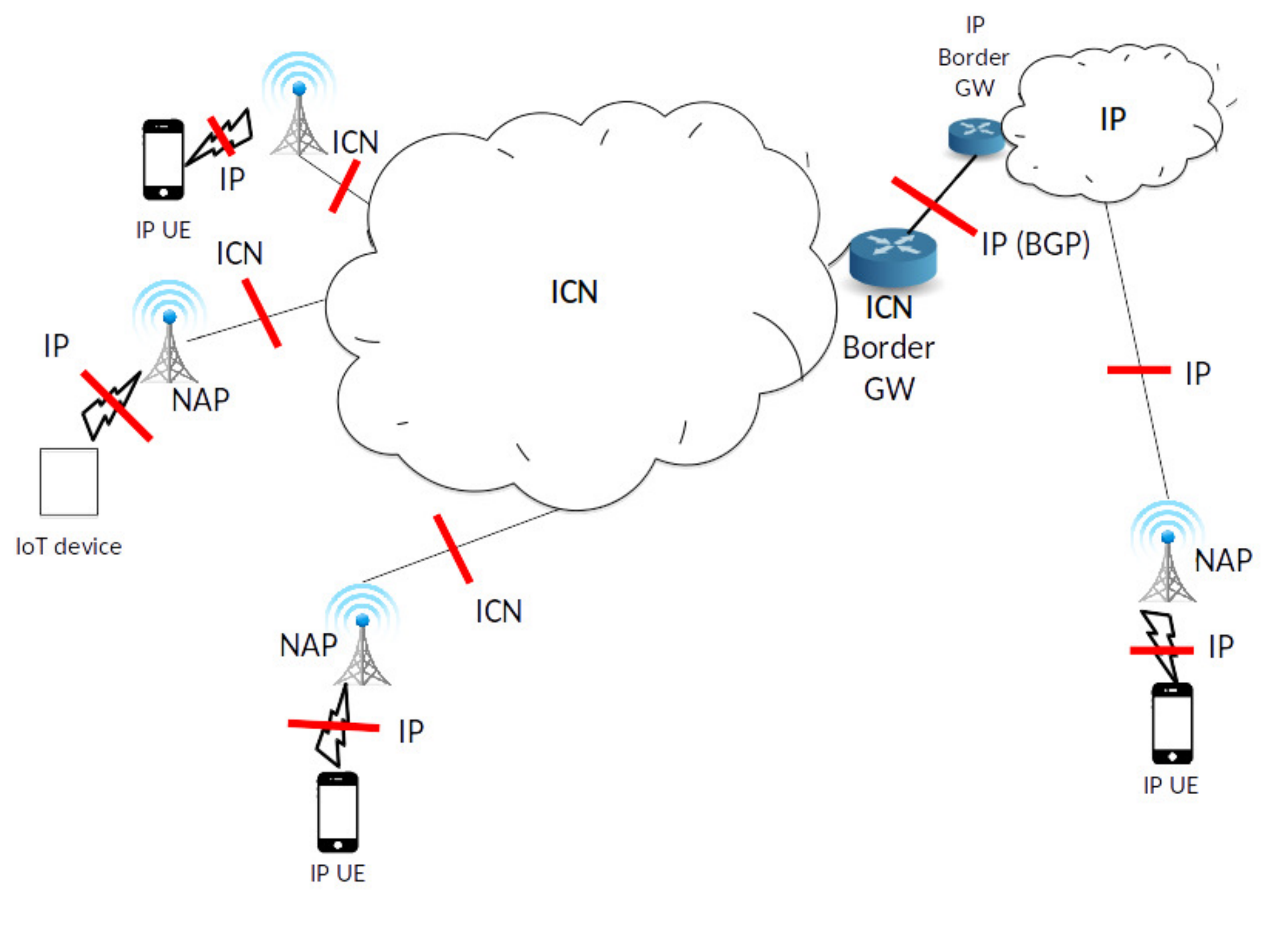}
\caption {The POINT architecture.}
\label{fig:point}
\end{figure}

\subsection{The POINT architecture}

Instead of dictating a clean-slate end-to-end ICN architecture, which would be very challenging to deploy, POINT allows standard IP traffic to be run over an ICN core network in a more efficient way~\cite{Tro2015}. To achieve this, the POINT architecture (Figure~\ref{fig:point}) provides a number of handlers implemented by the Network Attachment Points (NAPs). These handlers perform translations between the existing IP-based protocols (e.g., HTTP, CoAP, basic IP) and appropriate named objects within the ICN core on both edges of the core. Therefore existing applications can benefit from ICN's features such as native multicast and caching without any modifications. The potential benefits of the POINT architecture compared to an IP-based network, highlighted in more detail in~\cite{Tro2015}. 

In the POINT architecture, every content item is identified by a flat identifier known as  the Rendezvous  Identifier (RId).  Moreover,  every  content 
item belongs to at least one scope. The purpose of a scope is to group ``similar'' content items and to give a hint about content location.  Scopes  are  hierarchically organized and identified by a Scope Identifier
(SId). Scopes are managed by specialized Rendezvous Nodes (RNs), which form an  overlay Rendezvous  Network.  The  rendezvous  network provides a lookup service, which routes a ``subscription'' to a RN that ``knows'' (at least) one publisher for the requested item. A typical  ICN transaction  in  POINT  involves the following steps. A content item is assigned with
a RId and stored in (at least) one publisher that 
advertises its availability  in  one  or  more  scopes. With  this 
advertisement, the RId is stored in the RNs that manage these 
scopes.  Subscribers  send  subscriptions  for  specific   (SId,RId) 
pairs, which are routed by the rendezvous network 
to an appropriate RN. Upon receiving a subscription
message and provided that at least one publisher exists, the RN instructs 
a Topology  Manager to  create  a  forwarding  path  from  a 
publisher to the subscriber, which is included in the notification 
message to the publisher. Finally, the 
content item is transferred from the publisher to the subscriber.

\begin{figure*}
\centering
\includegraphics[width=.8\linewidth]{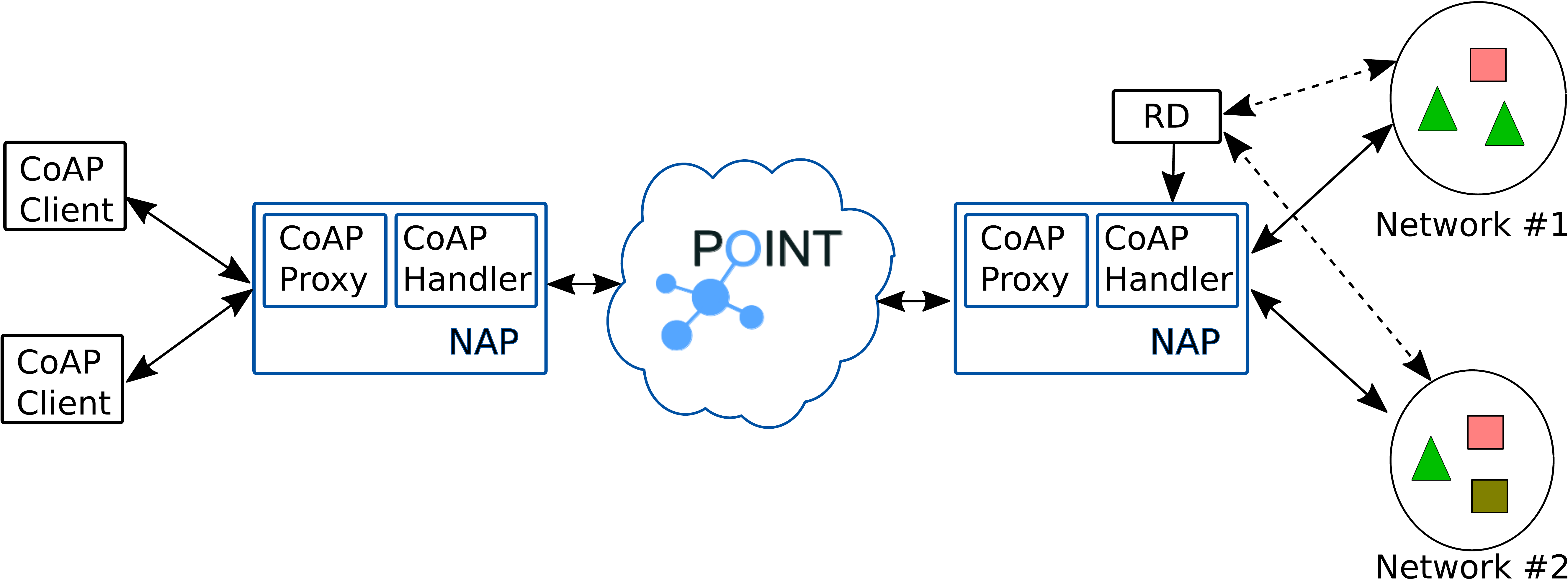}
\vspace*{.5cm}
\caption {An example of CoAP over ICN reference architecture. On the right part there are Things offering resources. Each resource is specified by a color and also by shape. On the left part there are CoAP clients.}
\label{fig:arch}
\end{figure*}

 \vspace*{.3cm} 
\section{CoAP Observe over ICN}
\label{sec:coap}

\subsection{Motivation}
\label{sec:motivation}

The IoT is expected to interconnect billions of heterogeneous devices, ranging from wireless sensors to actuators, wearable devices, Radio-Frequency Identification (RFID) tags, home appliances, surveillance cameras, and many others . It is expected that these devices will be uniquely identified and will be capable of communicating with each other.  However, the design of a networking architecture for the IoT, which is mainly composed of resource-constrained devices that generate highly heterogeneous traffic patterns, poses great challenges \cite{sheng2013survey}. First, there is the need for supporting resolution systems that translate resource URIs into IP addresses. However, it is difficult for constrained devices to allocate more resources for a DNS client implementation. Second, some of the constrained devices (e.g., temperature sensor) may receive vast amounts of requests  which requires significantly higher processing capabilities. Third, when CoAP observe is used a CoAP server needs to maintain state for each client and respond separately to each of them. 
 \vspace*{.3cm} 
\subsection{Design}
We now discuss the design of CoAP observe over ICN. This module is a part of our earlier CoAP over ICN work \cite{coaphandler} that illustrates various CoAP-specific communication scenarios.

Figure~\ref{fig:arch} illustrates the network setup for observing resources over the ICN architecture. In the middle of the figure there is the POINT network that interconnects NAPs. In the right part of the figure there are networks of \textit{Things}. Each Thing acts as a CoAP server offering a resource; the same (type of) resource can be offered by many Things located in different networks (e.g., there can be many sensors deployed in various parts of a city offering temperature measurements). Each network of Things is connected to the POINT network through a NAP. A network of Things may be directly attached to a NAP. A CoAP Resource Directory (RD) hosts the descriptions of resources provided by the CoAP servers. In the left part of the figure there are CoAP clients. A CoAP client is also connected to the POINT network though a NAP.

The fundamental component of our CoAP over ICN architecture is a \textit{CoAP handler} which is part of the NAP. A CoAP handler receives CoAP requests from CoAP clients (over IP), performs protocol translation and forwards the requests to CoAP server. The CoAP server generates a response which is forwarded through the ICN network to the CoAP clients following the reverse process.

\subsection{\textbf{Functional Requirements}}
To implement CoAP observe over ICN, a CoAP handler needs to satisfy the following functional requirements:

\vspace*{.2cm}

\begin{itemize}
 \item The CoAP protocol runs on top of UDP. A CoAP handler MUST maintain state of the CoAP clients prior forwarding their requests to the POINT network. This enables the CoAP handler to forward the corresponding response back to the appropriate client. \vspace*{.2cm}
 \item Efficient state maintenance within a CoAP handler in case of resource subscriptions. A CoAP handler should maintain additional state when similar requests are issued by multiple clients attached to the same client-side NAP (cNAP). Similarly, server-side NAP (sNAP) should also consider how to efficiently handle multiple requests for the same resource from multiple cNAPs.
 \vspace*{.2cm} 
 \item The CoAP handler MUST follow the protocol semantics of CoAP observe~\cite{rfc7641} when registering a CoAP client to the CoAP server. 
\end{itemize}

 \vspace*{.3cm} 
 
 \begin{figure}[!h]
\vspace*{.4cm}
\includegraphics[width=\linewidth]{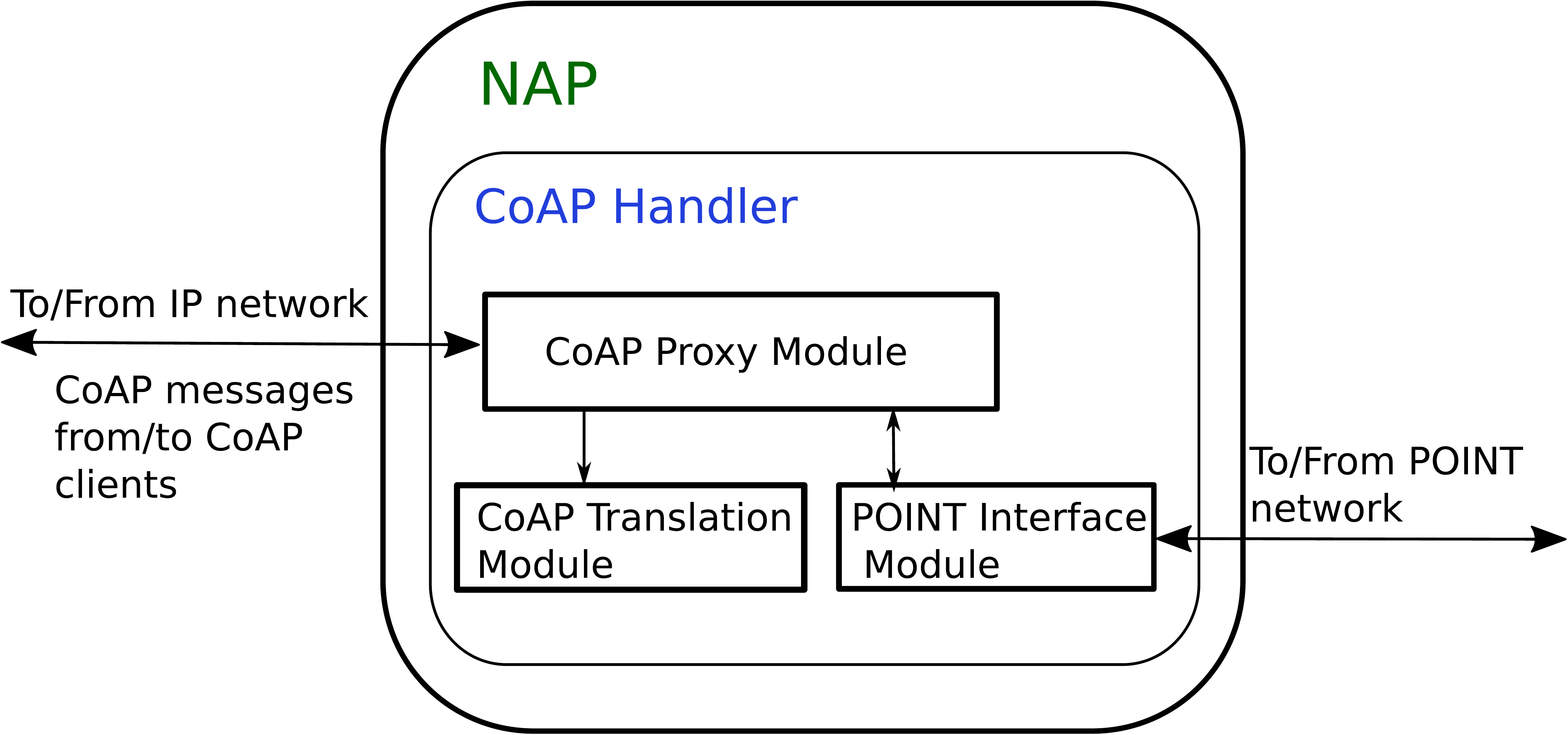}
\caption {CoAP handler functional module.}
\label{fig:coap-handler} \vspace*{.3cm}
\end{figure}

\subsection{\textbf{Functional Modules}}

A CoAP handler is composed of the following modules: \vspace*{.2cm} 

\textbf{Proxy:} A CoAP handler can be classified as one of the two categories depending on its role in the POINT architecture: forward proxy when it performs requests on behalf of the client, and reverse proxy when it behaves as if it were the original server. In our design, both roles are combined in the same module and are complementary. \vspace*{.2cm} 

\textbf{Protocol Translation:} This module allows the CoAP handler to translate CoAP messages to ICN messages and vice versa. The translation module follows the semantics of the CoAP protocol (RFC 7252). \vspace*{.2cm} 

\textbf{POINT Interface Module:} This module advertises ICN messages (i.e., translation of CoAP requests) to the ICN network. These messages eventually trigger the ICN rendezvous process, which leads to the forwarding of these messages to the appropriate NAP(s) (on the other side of the ICN network). A NAP that receives an ICN message restores the original CoAP request and forwards it to the appropriate CoAP server. The CoAP server generates a response which is forwarded through the ICN core network back to the CoAP clients following the reverse process.

\begin{algorithm}
\caption{Handling GET request for CoAP observe}
\label{alg:get}
\label{interest}
\begin{algorithmic}[1]

\STATE $isObserveOption \leftarrow [coap\_request]$
\STATE $token \leftarrow [coap\_request]$
\STATE $messageID \leftarrow [coap\_request]$
\STATE $resourceURI \leftarrow [coap\_request]$
\IF{$isObserveOption$ = = $ true $}
   \STATE $search\_subscription\_list(tokenFlag, uriFlag)$
   \IF{$tokenFlag$ = = $ true $}
     \STATE $ drop\ the\ request\ packet $ 
   \ENDIF  
    \IF{$uriFlag$ = = $ true\  \&\&\  tokenFlag \neq true $}
     \STATE $insert\ an\ entry\ in\ subscription\_list$ 
    \ELSE
      \STATE $create\ coap\ request\ with\ observe\ option $  
   \ENDIF  
 \ENDIF 
 \end{algorithmic}
\end{algorithm}

\section{\textbf{Node Operation}}
\label{sec:node}

\subsection{Module Overview}

The proxy module of the CoAP handler implements the core functionality of a CoAP proxy. The translation process of CoAP messages to ICN messages is implemented by this module. A CoAP request message and the information related to it are stored in a \texttt{client$\_$node} structure. Since CoAP transport is based on UDP, the \texttt{client$\_$node} structure allows a CoAP handler to match a response with the corresponding request and eventually to the appropriate CoAP client. The CoAP proxy module implements a list of \texttt{client$\_$node} to keep track of pending requests and resources observations. \\

\begin{verbatim}
struct client_node {
  struct client_node* next;
  struct sockaddr_storage addr;
  socklen_t addr_len;
  unsigned char * token;
};
\end{verbatim}

When a CoAP handler receives CoAP request messages the proxy module extracts the first 4-byte mandatory header, which contain the basic information of the request, including message type, method code (GET, PUT, DELETE, UPDATE), and token length. The URI of the resource, in a proxy request, is encoded as a string in the PROXY-URI option. A request may also include a Token which is used for matching requests and with (asynchronous) requests. The PROXY-URI can be split into the URI-HOST, URI-PORT, URI-PATH, and URI-QUERY fields. The URI-HOST is the FQDN of the CoAP server and the URI-PATH is the path of the resource within the server. 

The proxy of the sNAP constructs a new CoAP request; this request uses the 4-byte mandatory header of the original request and includes all the options extracted from PROXY-URI option. The new request also includes the Token of the original request.

\begin{algorithm}
\caption{Handling coap response from coap server}
\label{alg:response}
\begin{algorithmic}[1]
\STATE $isObserveOption \leftarrow [coap\_response]$
\STATE $token \leftarrow [coap\_response]$
\IF{$isObserveOption$ = = $ true $}
  \STATE $ found \leftarrow search\_observer(token, resource_uri) $
  \STATE $observer\_entry \leftarrow subscription\_list$
  \WHILE{$observer\_entry \neq NULL\  \&\&\  observer\_entry \rightarrow resourceURI$ == $resource\_uri $} 
      \STATE $token \leftarrow observer\_entry.token $
      \STATE $messageID \leftarrow observer\_entry.messageID) $
      \STATE $iteration \leftarrow observer\_entry.firstResponseFlag $	
      \IF{$iteration$} 
        \STATE $ \# To\ create\ observe\ relationship $
        \STATE $insert\ ACK\ code\ in\ response$
      \ENDIF
      \STATE $observer \leftarrow observer\_entry.client$
      \STATE $update\ response\ with\ token\ and\ messageID $
      \STATE $send\_response(coap\_response, observe)$
      \STATE $next\ observer\_entry $
   \ENDWHILE
 
 \ENDIF 
\end{algorithmic}
\end{algorithm}
\vspace*{.3cm}

\subsection{Handling Observe Request}
\label{sec:request}

Algorithm~\ref{alg:get} illustrates how an observe request in processed by the the CoAP handler. Processing observe requests is performed only if the request message contains the CoAP observe option. The first step of processing such a request is the establishment of an observe relationship between the CoAP client and the CoAP server. To achieve this, the proxy module maintains a subscription list the structure of which follows:

\begin{verbatim}
 struct coap_subscription{
     char* resource_uri;
     int resource_uri_len;
     unsigned char* token;
     int token_len;
     int iteration;
     uint16_t message_id;
     struct coap_subscription* next;
     struct client_node* client;
   };
\end{verbatim}

\normalsize

First, the module checks the \texttt{coap$\_$subscription} list to verify if the request is a retransmission; if yes it drops it. Second, the module verifies whether the request is intended for a resource which already exists in the \texttt{coap$\_$subscription} list. If this is true, the module inserts a new entry in the \texttt{coap$\_$subscription} list and drops the request, otherwise, it forwards it to the ICN network.

\subsection{Handling Observe Response}
\label{sec:response}

Algorithm~\ref{alg:response} illustrates how a CoAP handler process a response. First, the handler checks if the response contains the CoAP observe option: if this option is found, the proxy follows the Algorithm~\ref{alg:response} to process the response packet. A response packet only echoes the token and the message identifier of the original request and does not contain any information about the request URI. Therefore, the proxy module first checks the \texttt{coap$\_$subscription} list to find a match: if a match is found, the module extracts the appropriate resource URI. Based on the resource URI information, the proxy collects all ``observers'' and distributes the response to them. The \texttt{coap$\_$subscription} list also provides information if this is the first response. This is necessary in a case when multiple CoAP clients are interested in the same resource for which at least one observe relationship has already been established. For all subsequent observe relationships, the proxy checks the value of \texttt{iteration}: if this value is zero, the CoAP handler updates the message status code with acknowledgement, inserts the appropriate message id, and sends the response to every interested observer; then it changes the \texttt{iteration} value to 1. When a CoAP client receives a CoAP response corresponding to an observe request, it verifies the token and message id. If the verification is successful the observe relationship is established.

\begin{figure*}[t]
\centering
\includegraphics[width=.9\linewidth]{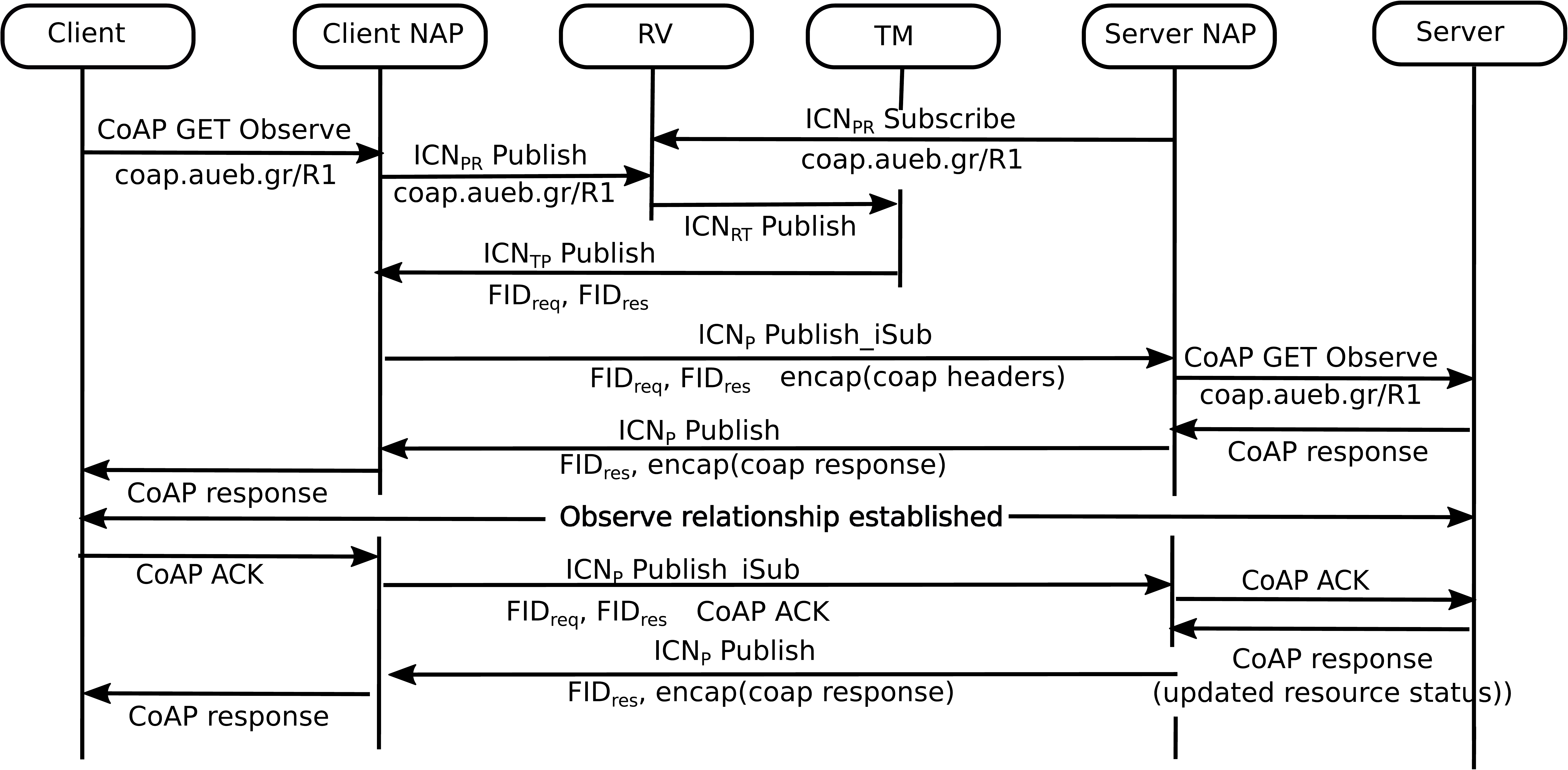}
\caption {Message Sequence Diagram from a CoAP client to CoAP server.}
\label{fig:msc}
\end{figure*}

 \vspace*{.3cm} 
\subsection{Handling Acknowledgement}
\label{sec:ack}

When a CoAP client issues an observe request, it waits for an acknowledgement from the CoAP server. With the reception of the acknowledgement, an observe relationship between the CoAP client and the CoAP server is established. The CoAP client also echoes a 4 byte mandatory header to notify the CoAP server that the CoAP client is alive and still interested in receiving further notifications. This procedure is intelligently handled by the CoAP handler in case of multiple observe requests for the same resource: the proxy module of the CoAP handler maintains the list of pending acknowledgements for those requests which are not forwarded to CoAP server. In more details, if a CoAP client sends an observe request to the CoAP server for a resource for which there already exists an observe relationship between the CoAP server and another client, the CoAP handler aggregates the request and suppress the acknowledgement. The CoAP handler acts as if it were the server of origin. The structure of the list is as follows:

\begin{verbatim}
 struct suppressed_token {
  unsigned char * token;
  int token_len;
  uint16_t message_id;
  struct suppressed_token* next;
};
\end{verbatim}

\begin{algorithm}
\caption{Handling ACK}
\label{alg:get}
\label{interest}
\begin{algorithmic}[1]
\STATE $message\_id \leftarrow [coap\_ack]$
\IF{$message\_id \in suppressed\_token$}
   \STATE $do\ nothing$
\ELSE
   \STATE $send\ coap\_ack\ to\ coap\ server$   
\ENDIF 
\end{algorithmic}
\end{algorithm}

\begin{figure*}
\centering
\subfloat[CoAP observe request.]{%
    \label{fig:req}
  \includegraphics[scale=.26]{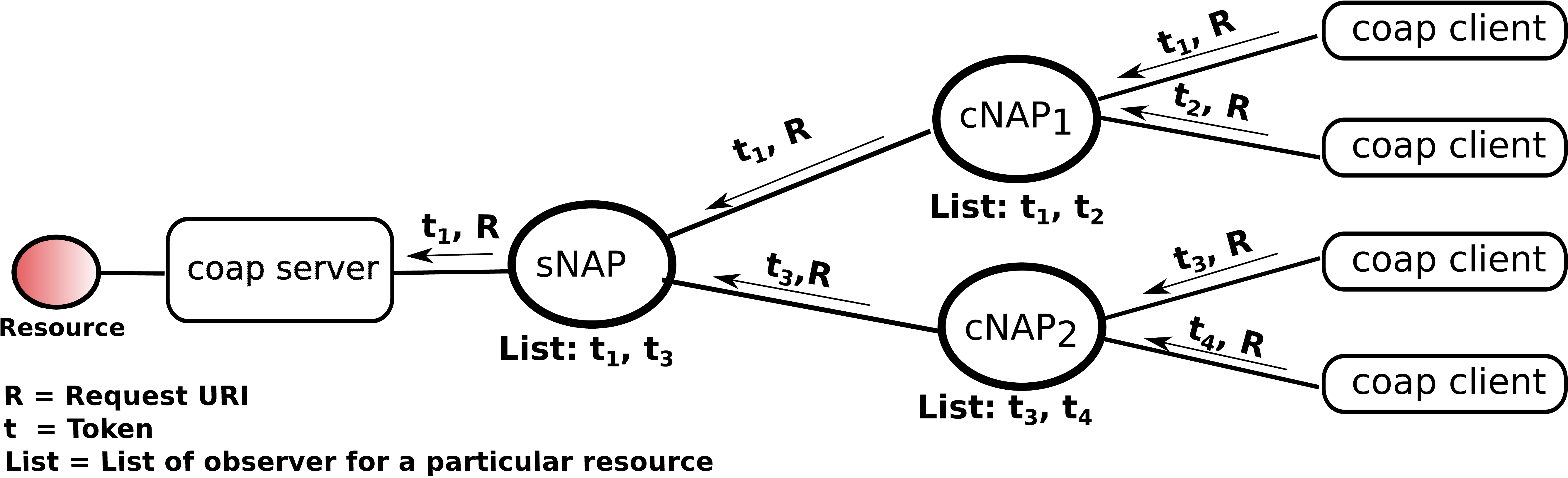}%
  }\par  \vspace*{.3cm}  
\subfloat[CoAP observe response.]{%
  \label{fig:res}
  \includegraphics[scale=.26]{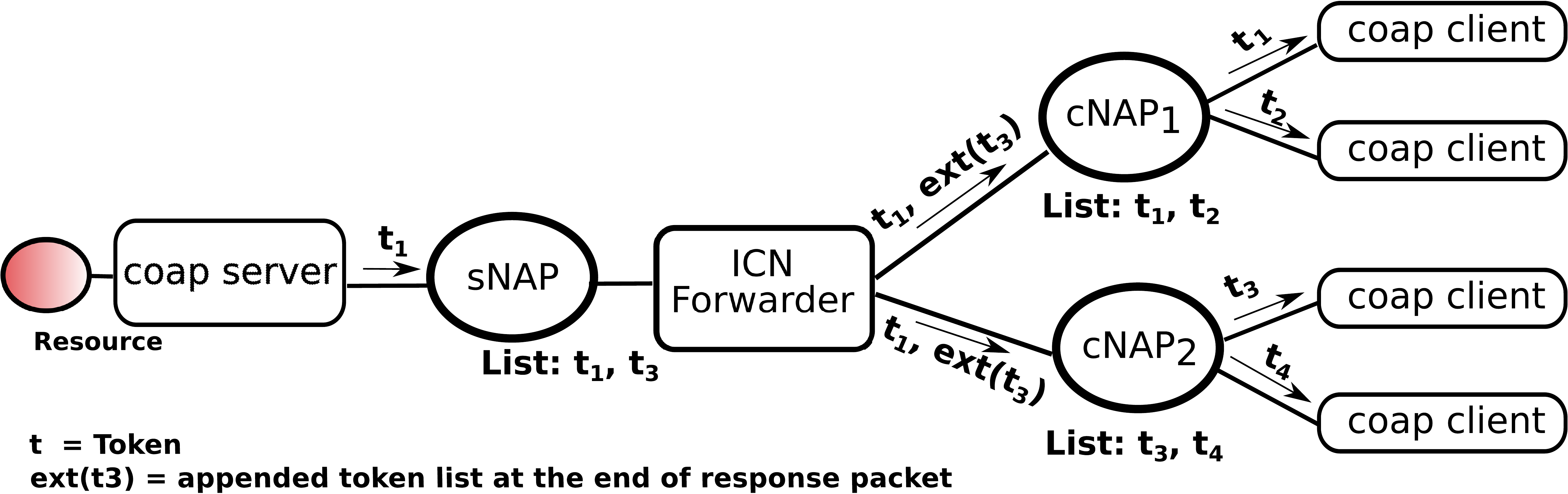}%
  }\par   \vspace*{.3cm}    
\caption{A simple scenario of observing a resource through POINT platform. (a) multiple observe requests for the similar resource \textbf{R} hosted in a CoAP server (b) Forwarding the response to CoAP clients. }
\label{fig:observe} \vspace*{.3cm}  
\end{figure*}

\subsection{ICN operations}
\label{sec:msc}

A request from a CoAP client is translated to an appropriate ICN name which is related to the FQDN of the server, while the response to that request is published to the appropriate ICN name related to the URL of the request. This allows a server NAP to simply subscribe to the FQDN of any attached CoAP server, while a CoAP server can publish any response to the corresponding URL.

In Figure~\ref{fig:msc}, we show an example of message sequence chart (MSC) for a CoAP client request and CoAP server response. The CoAP server registers its DNS name which leads to its NAP subscribing to the server's FQDN (aueb.example.gr/R1). CoAP client issues a CoAP observe request for the resource aueb.example.gr/R1. cNAP receives it and publishes to the Rendezvous (RVZ). the cNAP uses the pub$\_$isub API (publish with implicit subscription) call to publish the CoAP request. The RVZ matches it with the server’s subscription and asks the TM (as part of its internal realization) to create a forwarding path for the request as well as the reverse path. Eventually the TM sends the forwarding path (FID$_{req}$) and reverse path (FID$_{res}$) to the cNAP. cNAP forwards the request to sNAP using (FID$_{req}$). sNAP de-capsulate the CoAP request and sends it to the CoAP server. The CoAP server processes the request and sends the CoAP observe response to sNAP. sNAP forwards it to cNAP using (FID$_{res}$). Finally cNAP forward the response to appropriate client. Upon reception of the CoAP response, the observe relationship between the CoAP client and CoAP server is established. The subsequent communication from the CoAP client to transport ACK and the updated resource status is performed using (FID$_{req}$) and (FID$_{res}$).

 \vspace*{.3cm} 

\subsection{Evaluation}
\label{sec:evaluation}

We now present a simple CoAP communication scenario of observing a resource (Figure~\ref{fig:observe}) and discuss the benefits of the ICN to CoAP. In Figure~\ref{fig:req} CoAP clients issue observe requests for a resource \textbf{R}, including tokens \textbf{``t1", ``t2", ``t3", ``t4"} respectively. cNAP$_{1}$ receives CoAP observe requests from a CoAP client with  \textbf{``t1"}. cNAP$_{1}$ forwards the request to sNAP and creates an entry in the subscription list. sNAP receives the request and creates an entry in the subscription list for cNAP$_{1}$ and sends the request to the CoAP server. If the observe relationship is possible, the CoAP server replies with ACK which eventually creates an observe relationship between CoAP client and CoAP server through NAPs. After some time, another CoAP client issues observe request to  cNAP$_{1}$ with Token \textbf{``t2"}. cNAP$_{1}$ finds a match for this request and insert a new entry in the subscription list and does not forward the request to sNAP. Similarly, cNAP$_{2}$ receives two requests with  \textbf{``t3"}, \textbf{``t4"} and forward the request to sNAP with \textbf{``t3"}. sNAP receives the request from cNAP$_{2}$. sNAP finds a match and inserts a new entry for cNAP$_{2}$. cNAP and sNAP only forward the request which arrives earlier and maintain the list for others. The communication between cNAP and sNAP uses ICN message which is translated from the CoAP message.

The main benefit of the POINT architecture is its native multicast capabilities. Figure~\ref{fig:res} shows that CoAP server sends one unicast response to sNAP. sNAP translates it into an ICN message and sends it to cNAP$_{1}$ and cNAP$_{2}$. Finally cNAP$_{1}$ and cNAP$_{2}$ forward the response to the appropriate client including the correct tokens and message identifiers. In typical IP networks this communication pattern would result in multiple unicast transmissions from the CoAP server to the CoAP clients. In contrast, in POINT the impact to the network of this type of bursty traffic can be reduced by employing multicast. In order to achieve this, the CoAP handler instructs NAPs to use the same token for all these identical CoAP requests. NAPs are then responsible for modifying the token to the CoAP response.

Furthermore, by using ICN, we are able to provide better security and privacy for the constrained applications. Let us consider the example of a multi-tenant building, where various sensors have been
deployed. The building management system includes energy monitoring (e.g., temperature and humidity measurements), the security and safety of the building (e.g., motion detection, fire alarms), billing (e.g., energy consumption, number of parking slots used), and so on. It is expected that each tenant should be able to define access control policies since the information provided to building management system is very sensitive. However, it is not feasible to extend the constrained devices to support  access control policies, both from  performance/cost and security perspectives, as it will increase the processing power requirements and energy consumption. Nonetheless, a NAP is able to collect all information and implement information access policies. The CoAP handler is able to associate security and privacy requirements with namespaces, enabling the definition of fine-grained, reusable access rules that will govern information access directly from personal gateways.

As a proof of concept, we tested our design using two clients, one forward proxy, one reverse proxy and one CoAP server. The CoAP server and clients are implemented using libcoap\footnote{https://github.com/obgm/libcoap}. Forward and Reverse proxy runs our system. The CoAP clients issue observe requests for the same resource containing different tokens. The different tokens denote different CoAP request packets. The duration of observing the resource is 90s, which is the default timeout in the libcoap client implementation. We capture the CoAP traffic information using Wireshark. The CoAP traffic includes CoAP observe requests and CoAP acknowledgement packets. Our evaluation shows that only 48\% of the total request packets, including acknowledgments, are forwarded to the CoAP server, whereas the communication overhead in bytes is reduced by 50\%.

\section{Related Work}
\label{sec:related}

In recent years, several networking architectures have been proposed for Futrue Internet (e.g., CCN~\cite{jacobson2009networking}, DONA~\cite{koponen2007data}, PURSUIT/PSIRP~\cite{tarkoma2009publish}, POINT~\cite{Tro2015}, NetInf~\cite{dannewitz2009netinf} with aspiration to efficiently distribute and retrieve content. Various research efforts explore how these architectures can be used in the context of the IoT. At the same time, IPv6 over Low power Wireless Personal Area Networks (6LoWPAN) \cite{mulligan20076lowpan} is the current approach to connect Wireless Sensor Networks (WSNs) to the conventional Internet. In \cite{amadeo2014named}, the authors highlight the key challenges of IoT and provide a design of high level NDN architecture that can meet IoT challenges. In \cite{saadallah2012ccnx}, the authors propose the CCN communication layer on top of MAC layer to transmit packets. Similarly, an overlay of ICN architecture based on CCN on top of ETSI M2M architecture is presented in \cite{grieco2014architecting}. In \cite{cianci2012content}, authors propose a service platform based on CCN for Smart Cities that can integrate the available relevant wireless technologies to provide ubiquitous services, optimize the usage of communication resources through distributed caching and provide security by exploiting the security feature of CCN architecture. In \cite{baccelli2014information}, the authors provide an experimental comparison of CCN with the traditional IP based IoT standards 6LoWPAN/RPL/UDP in terms of energy consumption and memory footprint. This experiment has used a compact version of CCNx \cite{ccnx}, referred as CCN-Lite \cite{lite2014lightweight}, in RIOT OS \cite{baccelli2013riot}. The authors of \cite{franccois2013ccn} propose a push mechanism for CCN to optimize the traffic in sensor networks, whereas authors of \cite{song2013content} propose a content-centric internetworking scheme for resource constrained network devices based on task mapping where the network activities (e.g., storing, publishing, and retrieving content) of the constrained devices are transferred to the core CCN network. The work in \cite{6529776} has designed and implemented a lightweight CCN protocol targeted for Wireless Sensor Network as an alternative to IP protocol for sensor network. All these efforts require modifications to IoT endpoints. In contrast, the POINT architecture provides a CoAP handler that maps CoAP protocol onto appropriate named objects within the ICN core. 

Recent efforts \cite{ludovici2015proxy, correia2016dynamic} have been performed on proxy based CoAP observe in Wireless Sensor Network (WSNs). Alessandro et el. \cite{ludovici2015proxy} includes WebSocket protocol in the design of the CoAP proxy for HTTP based web applications. The work in \cite{correia2016dynamic} considers dynamic aggregation/scheduling of multiple observe requests at CoAP proxies. These efforts are complementary to our work, which utilizes ICN to further enhance the efficiency gains of the CoAP observe.

 \vspace*{.3cm} 

\section{Conclusions and Future work}
\label{sec:conc}

In this paper, we present the design and implementation details of CoAP observe for POINT architecture that enable CoAP clients to observe the resources hosted in IoT devices through ICN network. The CoAP clients are oblivious to the existence of ICN. CoAP observe functionality is very similar to publish/subscribe based ICN in particular asynchronous transmission. Transporting CoAP traffic over ICN can benefit in terms of communication overhead, state management and latency, in particular when multiple clients are interested to subscribe the same resource hosted in a IoT device. In addition, the inherent multicast capabilities of ICN and caching at the edge can be exploited in observing similar resources hosted in IoT devices by multiple CoAP clients. 

Future work in this are includes the integration of our solution in the POINT architecture. To evaluate the performance of CoAP observe over ICN, we will construct a simple IoT testbed which will be connected to the existing POINT testbed. 

 \vspace*{.3cm}

\section*{Acknowledgments}
Many of the ideas presented in this paper stem from discussions among POINT consortium partners. 
The work presented in this paper was supported by the EU funded H2020 ICT project POINT, under contract 643990.

 \vspace*{.3cm} 

\bibliographystyle{IEEEtran}
\bibliography{IEEEabrv,ntms}

\end{document}